\documentstyle[12pt,aaspp4,rotate,flushrt,epsf]{article}
\newcommand{\bl}{\large\bf}
\newcommand{\be}{\begin{equation}}
\newcommand{\ee}{\end{equation}}
\newcommand{\bc}{\begin{center}}
\newcommand{\ec}{\end{center}}
\newcommand{\ba}{\begin{array}}
\newcommand{\ea}{\end{array}}
\newcommand{\bea}{\begin{eqnarray}}
\newcommand{\eea}{\end{eqnarray}}
\newcommand{\beas}{\begin{eqnarray*}}
\newcommand{\eeas}{\end{eqnarray*}}
\newcommand{\et}{\end{tabular}}
\newcommand{\abtwo}{2 \sqrt{1-2\beta}\;\afa\frac{y_{\ell s}}{y_{os}}}
\newcommand{\acr}{a_{\rm cr}}
\newcommand{\afa}{\alpha_{0}}
\newcommand{\dtheta}{\Delta \theta}
\newcommand{\dthefa}{\frac{\dtheta}{\afa}}
\newcommand{\hnot}{{\rm H_{o}}}

\newcommand{\om}{\Omega_{\rm o}}
\newcommand{\omda}{\Omega_{\Lambda}}
\newcommand{\omr}{\Omega_{\rm R}}
\newcommand{\rc}{r_{\rm c}}
\newcommand{\rif}{r_{\infty}}
\newcommand{\rr}{\frac{\rif}{\rc}}
\newcommand{\sigtwo}{\sigma^{2}}
\newcommand{\sigrc}{\frac{(\sigma^*/km/sec)^{2}}{\rc^*/(h^{-1}Mpc)}}
\newcommand{\yy}{\frac{y_{\ell s}}{y_{os}}}
\begin{document}

\rightline{\bf CWRU-P14-98}
\rightline{\bf astro-ph/9802343}
\rightline{February 1998}
%\vskip 0.5in
\bc
%\vspace{0.5in}
%{\bf GRAVITATIONAL LENSING AND DARK GROUPS OR CLUSTERS}
\title{GRAVITATIONAL LENSING AND DARK STRUCTURES}
%\vskip 0.5in
\author{\sc Yu-Chung N. Cheng\altaffilmark{1}
and Lawrence M. Krauss\altaffilmark{2}}
\affil{Department of Physics,\\
Case Western Reserve University,\\
10900 Euclid Ave, Cleveland, OH 44106-7079}
\ec
\altaffiltext{1}{E-mail address: yxc16@po.cwru.edu}
\altaffiltext{2}{Also in Dept. of Astronomy, email:
krauss@theory1.phys.cwru.edu}

\begin{abstract}

\parindent=0.6cm

We examine whether a cosmologically
significant distribution of dark galaxy group
or cluster-sized objects can
have an optical depth for multiple imaging of 
distant background sources which is
comparable to that from known galaxies while at
the same time producing angular splittings of the same
order of magnitude.
Our purpose is to explore whether such objects could
realistically account for some of the observed lenses.  Modeling such 
systems as isothermal spheres with core radii, and assuming
a Schechter-type distribution function, we find that independent
of the cosmology (open, flat matter dominated, or flat cosmological constant
dominated) an allowed parameter range exists which is comparable in
velocity dispersion to that
for known compact groups of galaxies, although the preferred core
radii are somewhat smaller than that normally
assumed for compact groups.  Dark cluster-sized objects,
on the other hand, cannot reproduce the observed lensing
characteristics.  If the one known Dark cluster were
a good representative of such a distribution, most
such objects would not produce multiple images.
 We also present a result for
the angular splitting due to an isothermal sphere lens with non-zero core
radius, extending earlier work of Hinshaw and Krauss (1987). 
Our results are expressed as contour plots for
fixed lensing probabilities, and angular splittings.
\end{abstract}


\newpage
\setcounter{equation}{0}

\bc
\section{INTRODUCTION}
\ec

Ever since the first multiply imaged quasar was observed, it was clear
that the statistics of gravitational lensing could be utilized
to pin down cosmological parameters (i.e. \cite{TOG}; \cite{Ho}). Specifically,
for a given cosmological model, one can predict the optical depth
due to lensing by normal galaxies (presuming one has a model of
normal galaxies and their distribution), and compare that with 
observations.  Several large scale surveys have been performed, searching
for multiple imaging of distant quasars by intervening galaxies, and
more than a dozen such lensing events have been observed (\cite{kochanek96}).
However, there is one slight peculiarity.  While the overall frequency
of lensing events, and the rough angular splittings are reasonably
consistent with model expectations, in a significant fraction of the
cases, the actual lensing galaxy is not visible. Given the predicted
mass of the lensing systems, it is not obvious that such a large
fraction should remain unresolved. 

This prompts the natural question: Could the lenses be dark objects, perhaps
objects related to the distribution of dark matter in the Universe---perhaps
failed galaxies?
This is not a new idea, and it is one which has been
beset with problems.  In the first place,
anything close to a closure density of
compact objects generally produces an optical depth for lensing which is
too large (\cite{pressgunn}).  Second, if the dark objects are failed
galaxies, then their properties will generally preclude producing
multiple images with the observed angular splittings (\cite{HK}).

Here we examine another possibility.  Could larger systems, on the
scale of groups or clusters account for an observable fraction of the
known lenses?  The recent observation of a cluster-sized 
mass distribution containing one luminous galaxy (\cite{nature};
\cite{Mushotzky})
provides some additional {\it a posteriori} motivation for considering this
hypothesis.  

On first glance such
a possibility seems implausible.  Larger systems,
with larger velocity dispersions, will produce larger angular splittings, if they
produce multiple images at all.  Thus, it seems unlikely that such systems
might reproduce the observed lensing characteristics, which, as alluded to above,
are comparable to those one would predict for the known population of galaxies.
However, Hinshaw and Krauss (1987; hereafter HK) 
demonstrated that
under certain conditions a finite core radius suppresses the mean 
angular splitting due to isothermal sphere mass distributions. Here,
we generalize the earlier HK result, and prove that this
approximation is good for all lensing impact parameters inside the
critical disk for multiple image formation.  We then demonstrate that 
for a reasonable range of velocity dispersions, core radii, and total mass,
assuming a Schechter-type distribution function,
such systems can produce optical depths for lensing comparable to that due to
known galaxies, with comparable angular splittings.

It is well known that due to a variety of selection effects (magnification
biasing, etc.) the actual fraction of strong lensing events in any sample
can differ dramatically from the naive optical depth calculation.  However,
because these systematic effects should be largely independent of the
nature of the lenses themselves, if they are producing comparable angular
splittings, etc, we need not consider these effects in detail here.  
In particular,
we compare the calculated optical depth for lensing by dark 
objects to the naive optical depth for lensing by known galaxy distributions,
known to be in the range of $10^{-3}$ to $10^{-4}$.  Presumably, if the
other selection effects are comparable, this will then imply that such
systems could produce at least some fraction of the observed events.  We thus
derive contour plots in the parameter space of velocity dispersion and redshift
for fixed total optical depth in order to explore the suggested range of
dark lenses.  For this range we then explore the magnitude of the
predicted mean angular splittings.  We conclude with a brief summary
of our results.

\bc
\section{THEORETICAL FORMALISM}

\subsection{Optical Depth for Lensing}
\ec
%\par
We start with a mass density distribution with the
following form:
\be
\rho (r) = \frac{\sigtwo}{2 \pi G (r^{2}+\rc^{2})}
\label{eq:density}
\ee
where $\sigma$ is the velocity dispersion of this system, and $\rc$ is
the core radius. In order to
derive a finite total mass for the system 
we must assume some cutoff radius $\rif$. The total
mass is then given by:
\be
m(< \rif) = \frac{2 \sigtwo}{G} \left[ \rif - \rc \tan^{-1} \left( \frac{\rif}
{\rc} \right) \right]
\ee

If we assume a matter density fraction $\om$ in the Universe, 
then we can calculate the number density
of objects, $n$, as this fraction of
 the critical density $\rho_{\rm crit} = \frac
{3 \hnot^{2}}{8 \pi G}$ divided by the mass of each system:
\be
n = \om \frac{\rho_{\rm crit}}{m(< \rif)} = \om \frac{3 \hnot^{2}}{16 \pi 
\sigtwo \left[ \rif - \rc \tan^{-1} \left( \frac{\rif}{\rc} \right) \right]}
\label{eq:number}
\ee

Following Turner, Ostriker, \& Gott (1984)
we then define a lensing probability factor 
\be
F = \frac{c^{3} \pi n \afa^{2}}{\hnot^{3}} = \om \frac{3 \pi^{2} \sigtwo}
{c \hnot \left[ \rif - \rc \tan^{-1} \left( \frac{\rif}{\rc} \right) \right]}
\label{eq:ffactor}
\ee
Here, $\afa =
4 \pi (\frac{\sigma}{c})^{2}$, is the bend angle for
an isothermal sphere with core radius
$\rc = 0$.

Then, the differential optical depth is:
\be
d\tau = F \left[\frac{y_{o\ell}y_{\ell s}}{y_{os}} \right] ^2
[f(\beta_0)] \frac{dz_{o\ell}}{\sqrt{\om (1+z_{o\ell})^{3} + \omr
(1+z_{o\ell})^{2} + \omda}}
\label{eq:difp}
\ee
where $y_{oi}$ is the angular size distance (\cite{Peebles}):
%%%fraction of the proper distance compared to the horizon (\cite{Peebles}):
\be
y_{oi} \equiv \frac{\hnot a_{0} r_i}{c} =
\frac{1}{\sqrt{\omr}}\sinh\left(\sqrt{\omr}
\int_{0}^{z_i} \frac{dz}{\sqrt{\om (1+z)^{3} + \omr (1+z)^{2} + \omda}}\right)
\label{eq:ydist}
\ee
and
\be
f(\beta_0) \equiv 1 + 5\beta_0 - \frac{1}{2} \beta_0^2 - \frac{1}{2} \sqrt{\beta_0}
(\beta_0 + 4)^\frac{3}{2}
\ee
We will explain the origin
of the function $f(\beta_0)$ and the parameter $\beta_0$ in the next subsection.
In the above equations, 
$a_{0}$ is the scale factor of the Universe, $r_i$ is the coordinate distance, 
$z_i$ is the redshift, $y_{o\ell}$ is the
distance between the observer and the lensed galaxy, $y_{\ell s}$ is the 
distance between the lens and the source, and $y_{os}$ is the
distance between the observer and the source.
$\omr$ is the curvature term, and $\omda$ is the cosmological constant, so
that
$\om + \omr + \omda = 1$ in all cases.

In a flat universe, $y_{\ell s}$ is simply $y_{os}-y_{o\ell}$. However, in
an open universe (with $\omda =0$):
\be
y_{\ell s} = y_{os}\sqrt{1+\omr y_{o\ell}^{2}}-y_{o\ell}\sqrt{1+\omr y_{os}^{2}}
\ee

If all lensing systems had the same mass and core radii, this would then
be sufficient to calculate the relevant optical depths.  However, if one
has a distribution of masses and core radii, it is necessary
to integrate over this distribution.  For simplicity, we
assume that these systems are distributed, like observed galaxies and
clusters, with an effective Schechter function, with a parameter $L$,
which is related to the total mass of the system.
\be
\phi(L) dL = \left(\frac{L}{L^{*}}\right)^{\alpha_{1}} 
%\mbox{\Huge $e$}^{-\frac{L}{L^{*}}}
\exp \left(-\frac{L}{L^{*}}\right)
d\left(\frac{L}{L^{*}}\right)
\label{eq:schechter}
\ee
We can then consider $\sigma$ and $\rc$ as  
functions of $L$, and then integrate over this distribution when
deriving optical depths.
We use the conventional parameter $L$ to parametrize this distribution. 
However, because we are interested
in dark structures,  $L$ should be understood to refer to mass, rather
than luminosity in this case, unlike its representation for
luminous objects. Although one could easily choose to other
arbitrary mass distribution functions for
equation~(\ref{eq:schechter}), it seems reasonable to use the Schechter
function, since this fits the distribution of luminous galaxies.  Our chief
purpose here, in any case, is to explore what general mass ranges are
picked out in order to fit the data, and this rather general 
parametrization is as useful as any other for this purpose.

\bc
\subsection{Lensing Cross Section and Angular Splitting}
\ec

In order to
determine the cross section for strong lensing 
events, one needs to determine the
bend angle, $\alpha$, of the light trace from the source,
which in the case of an isothermal sphere
with a finite core radius, 
is a function of (\cite{HK}) the velocity dispersion
$\sigma$, impact parameter $b$, and core radius $\rc$.
The general formula for the bend angle is:
\be
\alpha (b) = \frac{4 b}{c^{2}} \int_{b}^{\infty} dr \frac{\partial \Phi}
{\partial r} \frac{1}{\sqrt{r^{2} - b^{2}}}
\ee
where
\be
\Phi (\vec{r}) = - \int_{v} d^{3}\vec{r^\prime} \frac{G \rho (\vec{r^\prime})}
{\left| \vec{r} - \vec{r^\prime} \right|}
\ee

HK derived a closed form approximation for
$\alpha$ using a mass distribution of the form.
Here, because there will be a relation between $\rif$ and the
number and mass density of dark objects, we have to refine
this earlier result to allow for finite $\rif$.  It
is simple to show that in this case, a first order expansion
in  $(\rc^{2}+b^{2})/ \rif^2$ leads to
\be
\alpha (b) = \afa \left( \frac{\sqrt{b^{2}+\rc^{2}} - \rc}{b} - \frac{b}{\pi 
\rif} \right) + o \left( \frac{b^{2}+\rc^{2}}{\rif^{2}} \right)
\ee
Following HK, we define $\bar{\lambda} \equiv
\acr / \afa$, where 
\be
\acr = 
\frac{c\,\afa\,y_{o\ell}\,y_{\ell s}}{\hnot (1+z_{\ell}) y_{os}}
\ee
so that
\be
\frac{b + l}{\bar{\lambda}} = \alpha (b)
\ee
where $l$ is the transverse distance of the lens center from the line of sight.

Defining
\bea
\bar{b}   & \equiv & \left( 1 + \frac{\acr}{\pi \rif} \right) b \\
\bar{\rc} & \equiv & \left( 1 + \frac{\acr}{\pi \rif} \right) \rc 
\eea
one can show
\be
\bar{b} + l = \acr \frac{\sqrt{\bar{b}^{2} + \bar{\rc}^{2}} - \bar{\rc}}
{\bar{b}}
\ee
This equation is identical in form to that derived by HK, so we
can then use their results to directly write down the
cross section
\be
\sigma_{\rm cs} = \pi \acr^{2} f(\beta_0)
\ee
where 
\be
\beta_0 = \frac{\bar{\rc}}{\acr} = \frac{\rc}{\acr} + \frac{\rc}{\pi
\rif} \leq \frac{1}{2}
\ee
Also note that $f(\beta_0)$ is zero when $\beta_0$ is larger
than $\frac{1}{2}$.

The bend angle not only allows us to derive the lensing cross section,
but also the observed angular splitting between lensed images. HK
demonstrated that for a lens located along the
line of sight to the source, this splitting is
given by $2 \sqrt{1-2\beta_0}\;\afa\frac{y_{\ell s}}{y_{os}}$.  In an appendix 
we present an analytic proof that this approximation
is good to $10 \%$ for all values of $l$ relevant for multiple
image formation.  Hence we include this factor in our analysis. Note
that, depending on the value of $\beta_0$, this can lead to
significant reductions in the mean predicted splitting, allowing
systems with velocity dispersions larger than those of galaxies to produce 
comparable angular splittings.

\bc
\section{NUMERICAL RESULTS}
\ec

Utilizing the formalism developed in the last section, we can
estimate the optical depth due to dark objects of specified
velocity dispersion and core radius for several different cosmologies.
We consider first the results obtained without averaging over
a distribution of objects, and then explore how such averaging
can impact upon the optical depths obtained.  We then focus on
considerations of the mean angular splitting produced by such
dark lenses, and finally make some comparisons to observations.

Figure 1 displays contour plots for optical depths of $10^{-3}$ (thick
lines) and $10^{-4}$ (thin lines) with no averaging over a
distribution of masses, but instead taking a uniform cosmic density
of objects of fixed mass, leading to net density $\Omega_0$.  Note that
there are three free parameters, $\sigma$, $\rc$ and $\rif$.  For
a fixed ratio $\rr$, related to the mass per object, the
integrated optical depth out to source redshift $z_s$
is then simply a function of $\sigma^2 / \rc$, so it seems reasonable
to present
the results in the phase space for this ratio, normalized
in units of $\frac {(km/sec)^2}{h^{-1}Mpc}$ vs $z_s$.
Comparison to existing, or proposed, mass distributions can be
done by plugging in physical values for $\sigma$, $\rc$, or both.
Finally, all the black curves correspond to $\rr=10$, and all gray
curves correspond to $\rr=100$. The solid curve sets are for a flat universe,
with $\om=1$, the dash-dash curve sets are for a flat universe model
with $\om=0.3$, and $\omda=0.7$, and the 
dash-dot curve sets are for an open universe
model with $\om=0.3$ and zero cosmological constant.

Note first that the $\rr=100$ curve is higher
than the $\rr=10$ curve for a fixed cosmological model.
This can be understood as follows.  For a fixed value of $\rc$ and
$\sigma$, increasing $\rr$ increases the mass per system.  However,
since the overall number density of systems is determined by the
requirement of some fixed $\om$, this number density then decreases
inversely with $\rr$, leading to fewer lenses.  Since the lensing probability
for isothermal spheres (with core radius) is {\it not} determined
by their total mass, but rather by their velocity dispersion (and $\rc$),
having fewer lenses means a smaller optical depth, even though the mass
per lens increases.   Next note that while it is known that for
a fixed density of lenses, the optical depth increases for an open
universe model compared to a flat one $\om=1$, this effect is not seen here 
simply
because we normalize the number density by the matter density.  Since
an open universe has a lower matter density than a flat matter dominated
universe, the number density of lenses decreases in proportion, and
this is reflected in the optical depth contour plots shown here.  The
well known fact that the optical depth for a cosmological constant dominated
flat universe is larger than for a matter dominated flat universe is also 
reflected in our results.

Before proceeding to compare with observational data to determine
if these estimates might correspond to reasonable structures, we need
to allow for the fact that the dark lenses are distributed over different
masses as per equation~(\ref{eq:schechter}).

Without strong theoretical 
guidance in this regard, we choose, by analogy with luminous
objects, 
$\alpha_{1} = -1$ in equation~(\ref{eq:schechter}).
 We also assume $\frac{\rc}{\sigtwo}$ is proportional to
$L^{\alpha_{2}}$, where we choose $\alpha_{2}$ to be {\it
negative} to avoid a singularity. We
display our results for $\alpha_{2} = -0.5$ and $ -1.5$. 
Note that in their lensing
probability analysis for {\it luminous} objects, Krauss
and White (1992) suggested that  $\alpha_{2}$ 
could be a small positive number, although not much
larger than zero at best. 
Note also, however that if one assumes the core radius
is not a function of mass, then $\alpha_{2}$ is negative.
In Figs.~2 and 3 our results for the two different choices of $\alpha_2$
are shown. As can be seen, the curves move toward smaller values
of $\sigrc$ when $\alpha_{2}$ decreases. That is, the probability is higher
for a given source redshift and $\sigrc$, when $\alpha_{2}$ decreases, as
one might expect, since this implies that larger core radii are less heavily
weighted. We should remind our readers here that $\sigma^*$ and $\rc^*$ are
normalized to $L^*$ in equation~(\ref{eq:schechter}).

We next consider the mean
angular splitting induced by the lensing distribution. Again, we first
consider the case when we do not integrate over a mass distribution
of lensing.  Recall that for a singular isothermal sphere
distribution, the mean angular splitting is $\afa$.   In the case of finite
core radius,
the expectation value of $\dthefa$, using our result that
 $\dtheta  \approx 2 \sqrt{1-2\beta_0}\;\afa\;\yy$, is:
\bea
<\!x\!> \equiv \frac{1}{\tau} \int d\tau \,x \nonumber\\
<\!\dthefa\!> = \frac{1}{\tau} \int d\tau \dthefa
\eea
where $\tau$ is the optical depth from equation~(\ref{eq:difp}). 
In Fig.~4, we show
contour plots
for the case
when the expectation values of the  angular splitting either 0.5 or 0.8
times that expected for the equivalent singular isothermal sphere distribution. 
The thicker curves represent the expectation value of 0.8$\afa$, and the
thinner curves represent the expectation values of 0.5$\afa$. Other symbols
are the same as in Fig.~1. The vertical axis is
still $\frac{(\sigma/km/sec)^{2}}{\rc/(h^{-1}Mpc)}$ and the horizontal axis is 
source redshift. 

Note that
the $\rr=10$ curve is higher than the $\rr=100$ curve.  This is
opposite to
the probability plots discussed in the earlier subsections, and
reflects the simple fact that larger $\rc$ for a fixed $\sigma$ results
in a smaller angular splitting.   This tension between angular splitting
and optical depth is important, because it implies that to keep angular
splittings small enough to be comparable to those observed, one cannot
allow too large a value of $\sigma$ and still keep the overall optical
depth comparable to that due to luminous galaxies.  This tuning is
an important feature which constrains this scenario.

Finally, we repeat the above analysis, including an integration
over a Schechter distribution of lenses, as described above.  The parameters
are
the same as discussed above. In Fig.~5, we chooose
$\alpha_{2} = -0.5$, and the thick and thin curves
represent when the expectation values of the angular splitting are
respectively
0.9 and 0.7 of the zero core radius case. 
In Fig.~6, we choose $\alpha_{2} = -1.5$ and the thick and thin curves
represent when the expectation values of the angular splitting are 0.85 and 
0.95 of the zero core radius case. Note that
when $\alpha_{2}$  decreases, the expectation values for the
angular splitting approach the isothermal sphere result, as one might expect,
since for these values, the core radius is becoming less significant.
Thus, if $\alpha_{2}$ is less negative than we have assumed here, the
expectation value of the angular splitting could even be smaller in
comparison to the equivalent isothermal sphere distribution than we have
found.

\bc
\section{COMPARISON WITH OBSERVATIONS AND CONCLUSIONS}
\ec

Over a dozen multiply imaged quasars have now been observed in various 
optical and radio surveys.  While these have angular splittings
characteristic of those one might expect to be induced by 
galaxies (i.e. $O(2''$-$3'')$), 
several have no observable lensing systems, even when one
might expect the lensing galaxy to be resolvable.  When one calculates
the optical depth for lensing by known galaxy distributions, ignoring 
magnification biasing, one typically finds $ 10^{-4} \le \tau \le 10^{-3}$.
 Thus, one might expect that a predicted optical depth of as low as $10^{-5}$
for some other distribution would produce at least one or two lenses
in the existing surveys.  In fitting the optical depth, however, one
must confront the tension between optical depth and angular splitting discussed
above (also compare Fig.~1 and Fig.~4).
Systems with velocity dispersion greater than about $450 km/sec$ would
require the expectation value of the angular splitting to be less than
half that predicted for an equivalent singular isothermal sphere.  This,
in turn, requires $\rc$ to be larger relative to $\sigma$, which in turn, 
however, suppresses the optical depth.  Nevertheless, in Table 1 we display 
several sample values of core radii and velocity dispersions which would 
be expected to produce
a fraction of the observed lenses with angular splittings comparable to
those observed. In this table we have fixed $\rr =10$, set
the source redshift to 3, and fixed the parameters
so that the mean predicted angular splitting
is $2.5$ arcsec. Note that the parameter range is comparable to
compact groups of galaxies, which have
compatible $\frac{(\sigma/km/sec)^{2}}{\rc/(h^{-1}Mpc)}$. For example, one
study suggests $\sigma$ is around 331 $km/sec$, and 
$\rc$ is around $15 h^{-1} kpc$ (\cite{mog}; \cite{PBE}). These numbers give 
$\frac{(\sigma/km/sec)^{2}}{\rc/(h^{-1}Mpc)} \approx 7.3*10^{6}$, as 
compared to the preferred values of 
$\frac{(\sigma/km/sec)^{2}}{\rc/(h^{-1}Mpc)}$ 
between $10^{7}$ and $10^{9}$ to result in an optical depth comparable to
that of the known galaxy distribution.  Also note that if we
were to integrate over a distribution of such objects, our earlier
arguments suggest that comparable optical depths and splittings could
be obtained even if the mean value of $r_c$ were somewhat larger than
given in the table.

However cluster-sized objects do not seem to be viable candidates for dark 
lenses because of the larger predicted splittings when multiple image formation
does occur, and more importantly because the predicted optical depth is too 
small. (The $\rc$ of a regular cluster 
is large, but the velocity dispersion is only about 2.5 times larger than
that of compact groups of galaxies (\cite{zghv}). )  This
conclusion is reinforced by the recent observation of an actual dark cluster
by Hattori et. al. (1997) in the lensing system, MG2016+112. Based on the 
inferred mass of the object using X-Ray estimates of the potential,
its size, and the size of the core radius of this system, we find
that $\beta_0 > 1/2$, implying that the cluster is not 
responsible for the observed multiple images in this system. This is
again supported by the small angular splitting of $3.4$ arcsec between
the images, which is characteristic of the one observed galaxy in this system.  

Thus, perhaps paradoxically, the only known 
example of a large-scale dark object suggests that such cluster-scale
objects, even if they have a significant mass density in
the universe, are probably largely irrelevant for the statistics of multiply
imaging distant quasars. Rather, dark  objects on intermediate scales,
between galaxy and cluster scales, are more likely possible candidates for
dark lenses. A very recent paper which has shown a very high rate
of galaxy lensing in radio surveys may be significant in this 
regard (\cite{Helbig}). We also note both that
a recent study of the luminosity function of the compact
groups of galaxies also lends support to our
choice of Schechter $\alpha_{1} = -1$ (\cite{ZCR}),
and that the currently favored cosmological model involving
a flat universe with cosmological constant (i.e. \cite{krauss}) produces
the largest optical depths at high redshift for given 
$\frac{(\sigma/km/sec)^{2}}{\rc/(h^{-1}Mpc)}$, as one would expect, based on 
the increase in the optical depth for lensing by galaxies in this cosmology.

Clearly, in order to know whether dark clusters are important for
lensing in the actual universe, larger surveys will be required, in order
to reliably determine how many lensing events might not be associated
with galaxies.  If such events continue to be observed, our analysis
suggests that the distribution of angular splittings will be
an important observable which might constrain possible models.

\bc
{\bl ACKNOWLEDGMENTS}
\ec

\acknowledgments{
This research work has been partially supported by the Industrial Physics
Group in the Physics Department at Case Western Reserve University, and by
a grant from the  DOE.}

\newpage
%\bc
%{\bf APPENDIX}
%\ec
\appendix\section{APPENDIX}

We prove here the angular splitting, $\dtheta$, is
a monotonically decreased function of $L \equiv \frac{l}{\acr}$ (not to be 
confused with the quantity $L$ in the Schechter distribution function 
used in the text), where $l$ is the distance from the line of sight of
the lens.  Then, using this 
fact, we can show that $\dtheta$ is well approximated by $\abtwo$, the angular
splitting derived by Hinshaw and Krauss for lenses along the line of sight.
We start by redefining the parameters of equation~(14)
in Hinshaw \& Krauss (1987):
\be
x \equiv \frac{b}{\acr} \hspace{0.3in}
\beta \equiv \frac{\rc}{\acr} \hspace{0.3in}
L \equiv \frac{l}{\acr}
\ee
We then find
\be
x^{3} + 2 L x^{2} + (L^{2} + 2 \beta -1) x + 2 L \beta = 0
\ee
(We remind the reader that $\beta$ is between 0 and 
$\frac{1}{2}$, and $|L| \leq 1$.) The above algebraic equation is solvable.
Setting $x = y - \frac{2 L}{3}$
the solution is
\be
y = u+v, \hspace{0.3in}
-\frac{1}{2}(u+v)+i\frac{\sqrt{3}}{2}(u-v), \hspace{0.3in}
-\frac{1}{2}(u+v)-i\frac{\sqrt{3}}{2}(u-v)
\ee
with
\be
u = \sqrt[3]{-\frac{q}{2} + \sqrt{\left(\frac{q}{2}\right)^{2}+\left(\frac{p}
{3}\right)^{3}}}, \hspace{0.5in}
v = \sqrt[3]{-\frac{q}{2} - \sqrt{\left(\frac{q}{2}\right)^{2}+\left(\frac{p}
{3}\right)^{3}}}
\ee
\bea
-p          & \equiv & 1 - 2 \beta + \frac{1}{3}L^{2} \geq 0 \\
\frac{q}{2} & \equiv & \frac{L}{3}\left( 1 + \beta -\frac{1}{9}L^{2} \right)
\geq 0
\eea
If we want to have three real solutions of $y$, which correspond to three
lensing images, then $u - v$ should be a pure imaginary number. However,
$u + v$ is a real number, and this gives us $u^{*} = v$, and
$\sqrt{\left(\frac{q}{2}\right)^{2}+\left(\frac{p}{3}\right)^{3}}$
is a pure imaginary number. We thus define
\be
-r^{2} \equiv \left(\frac{q}{2}\right)^{2}+\left(\frac{p}{3}\right)^{3} \leq 0
\ee
\be
r^{2} = \frac{1}{27}[(1-2\beta)^{3}-L^{2}(2+10\beta-\beta^{2})+L^{4}] \geq 0
\ee
and 
\be
r=0 \Leftrightarrow L_{0}^{2} = \frac{l_{0}^{2}}{\acr^{2}} = f(\beta)
\ee
so we can re-write $u$ and $v$ as following:
\be
u = -\sqrt[6]{\left(\frac{q}{2}\right)^{2}+r^{2}}\;\exp(-i\phi/3), 
\hspace{0.5in}
v = -\sqrt[6]{\left(\frac{q}{2}\right)^{2}+r^{2}}\;\exp(i\phi/3)
\ee
where
\be
\tan\phi = \frac{2 r}{q} \geq 0
\ee
this implies that
\be
0 \leq \phi \leq \frac{\pi}{2}
\ee
Recall that $r^{2}+(\frac{q}{2})^{2} = (\frac{-p}{3})^{3}$, then we have
\be
u = -\sqrt{\frac{-p}{3}}\,\exp(-i\phi/3), \hspace{0.5in}
v = -\sqrt{\frac{-p}{3}}\,\exp(i\phi/3)
\ee
we can write down the three solutions of $y$:
\be
\ba{ccccl}
y_{1} & = & u + v        & = & -2 \sqrt{\frac{-p}{3}} \cos \left( 
\frac{\phi}{3}\right) < 0 \nonumber\\
y_{2} & = & -\frac{1}{2}(u+v)+i\frac{\sqrt{3}}{2}(u-v) & = & 
2 \sqrt{\frac{-p}{3}} \cos \left( \frac{\pi}{3}+\frac{\phi}{3}\right) \geq 0
\nonumber\\
y_{3} & = & -\frac{1}{2}(u+v)-i\frac{\sqrt{3}}{2}(u-v) & = &
2 \sqrt{\frac{-p}{3}} \cos \left( \frac{\pi}{3}-\frac{\phi}{3}\right) > y_{2}
\ea
\ee
therefore, the angular splitting is:
\be
\frac{\dtheta}{\yy} =\afa (\Delta (x+L)) = \afa \Delta y = \afa (y_{3}-y_{1})=
2 \afa \sqrt{-p} \cos \left( \frac{\pi}{6} - \frac{\phi}{3} \right) > 0
\ee

Based on the above equation, one can derive (after some
work) that:

\bea
\frac{1}{\afa\,\yy} \frac{d^{2}(\dtheta)}{dL^{2}} & = & \frac{2}{3} \frac{1}
{(\sqrt{-p})^{3}} \cos\left(\frac{\pi}{6}-\frac{\phi}{3}\right)\left[(1-2\beta)-
\frac{1}{3}(-p)^2 \left(\frac{d\phi}{dL}\right)^{2}\right] \nonumber\\
 & & \mbox{}
+\frac{2}{3}\sin\left(\frac{\pi}{6}-\frac{\phi}{3}\right)\left[\frac{L}
{3\sqrt{-p}} 
\frac{d\phi}{dL} + \frac{d}{dL} \left(\sqrt{-p}\frac{d\phi}{dL}\right)\right]
\leq 0
\eea

Note that $\dtheta$ varies from $L=0$ to $L=L_0$ :
\bea
\dtheta & = & \abtwo, \hspace{0.3in}{\rm when}\;L = 0\nonumber\\
\dtheta & = & \sqrt{3} \sqrt{1-2\beta+\frac{1}{3}L_{0}^{2}}\;\afa\;\yy, 
\hspace{0.3in}{\rm when}\;L = L_{0}
\eea

Finally, the average ratio of $\dtheta$ to $\abtwo$ at $L = L_{0}$ is
\be
2 \int_{0}^{\frac{1}{2}} d\beta \frac{\dtheta}{\abtwo}
\simeq 0.88
\ee

Since we have shown that $\dtheta$ is a monotonically decreasing function of
$L$, this implies that over the relevant range of $L$, 
$\dtheta$ does not differ by more than $10\%$ from its value at $L=0$.

\newpage
\begin{table}[t]
\bc
\section*{TABLE 1}
\subsection*{\sc Sample Values of Core Radii and Velocity Dispersion Relevant
to Lensing}
%Which Yield Reasonable Optical Depths and Angular Splittings}

\begin{tabular}{ccccc} \tableline\tableline
model & $\sigma(km/sec)$ & $\rc(h^{-1}kpc)$ & $\tau$ & $<\dtheta>$\\
\tableline
$\om=1, \omda=0$     & 427 & 4.6 & $2.9*10^{-5}$ & 0.47$\afa$\\
$\om=0.3, \omda=0.7$ & 369 & 5.0 & $6.9*10^{-5}$ & 0.55$\afa$\\
$\om=0.3, \omr=0.7$  & 416 & 4.4 & $3.4*10^{-5}$ & 0.50$\afa$\\ \tableline
\end{tabular}
\tablecomments{The values are obtained by assuming $\rr=10$, source redshift
at 3, and $<\dtheta>$ is fixed at 2''.5.}
\ec
\end{table}
\clearpage

\newpage
\bc\
\epsfxsize=5in
\epsffile{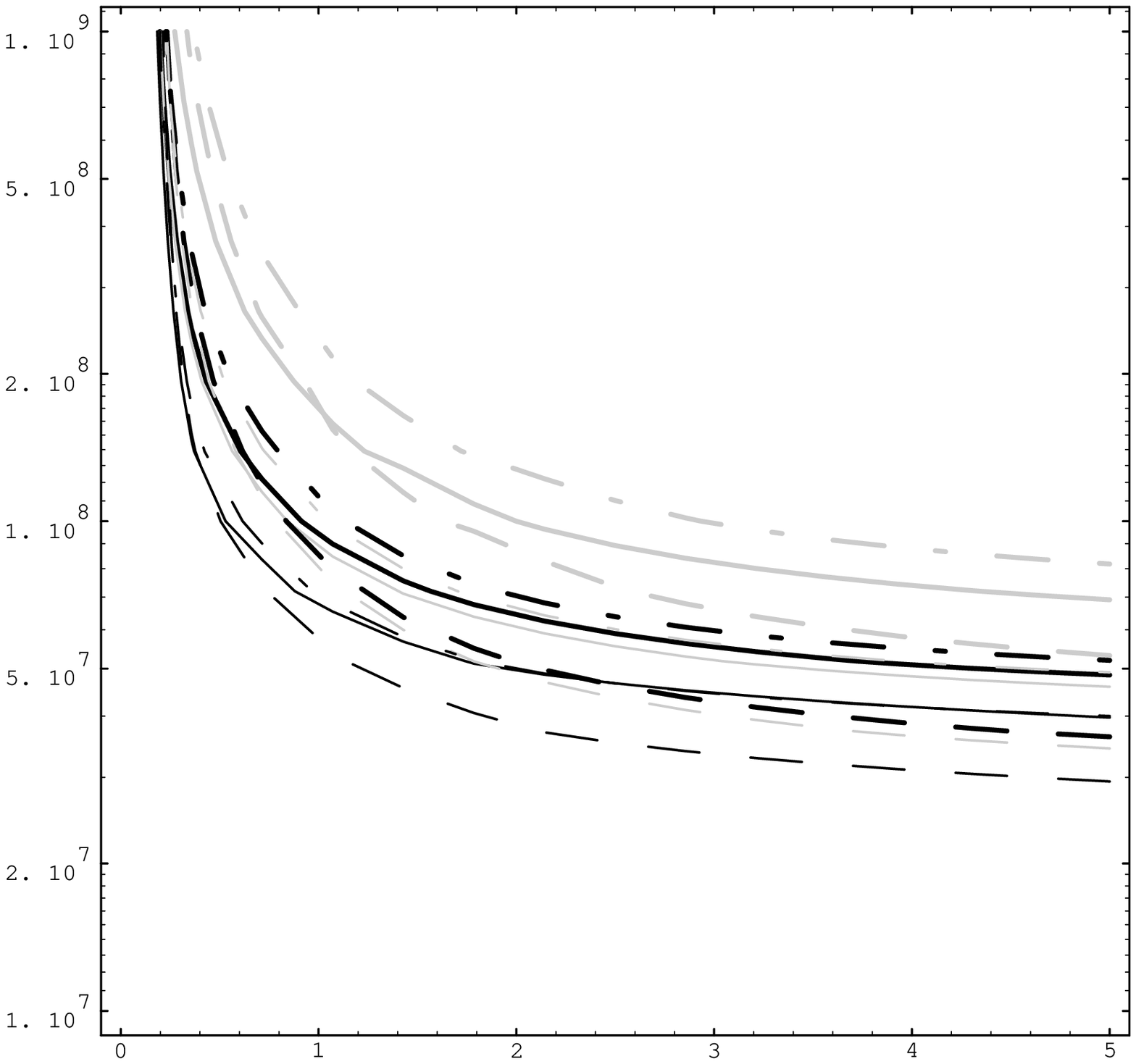}
\ec
\vskip -5.0in
{\hskip 3.05in \epsfxsize=2.9in \epsffile{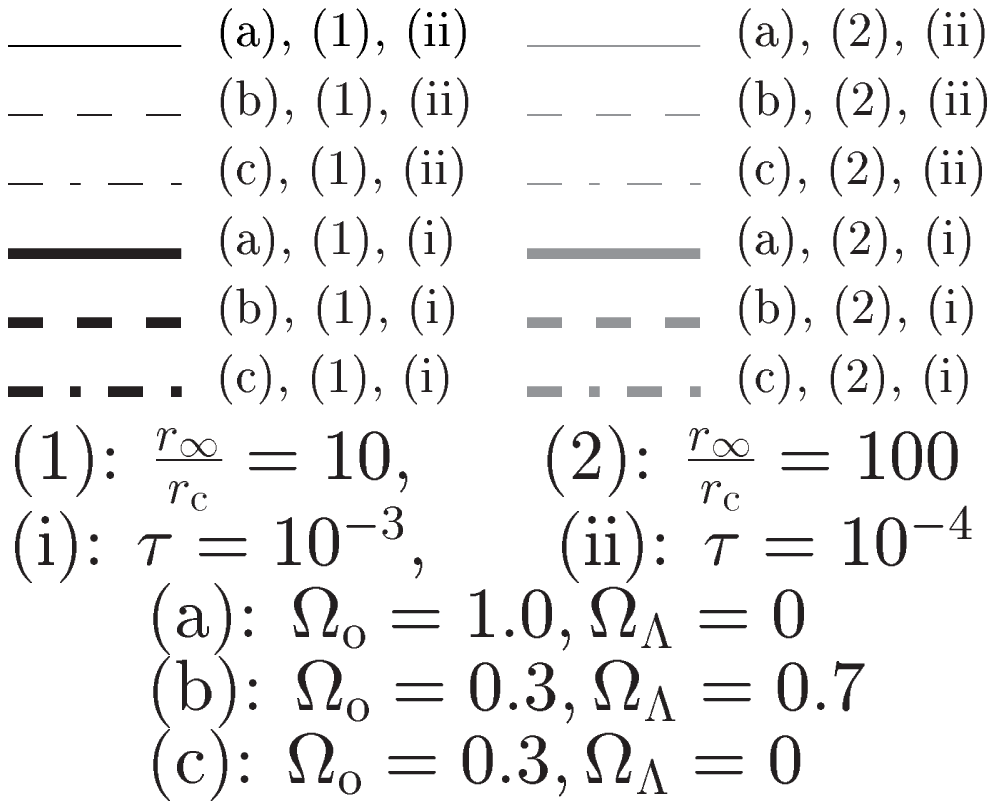}}
\vskip -0.3in
{\hskip -0.1in \Large \rotate[l]{$\frac{(\sigma/km/sec)^{2}}{\rc/(h^{-1}Mpc)}$}}
\vskip 1.8in
{\hskip 3.1in \Large $z_{s}$}
\vskip 0.2in

\figcaption{Contour plots of lensing probability for fixed lens mass. The 
thicker width lines represent an optical depth of 10$^{-3}$, and the thinner 
width lines represent an optical depth of 10$^{-4}$. The black curves 
correspond to $\rr=10$, and the gray curves correspond to $\rr=100$. The solid 
curve sets are for a flat universe model, with $\om=1$. The dash-dash curve sets
are also for a flat universe model, but with $\om=0.3$, and $\omda=0.7$. The 
dash-dot curve sets are for an open universe
model, with $\om=0.3$ and zero cosmological constant.}

\newpage
\bc\
\epsfxsize=5in
\epsffile{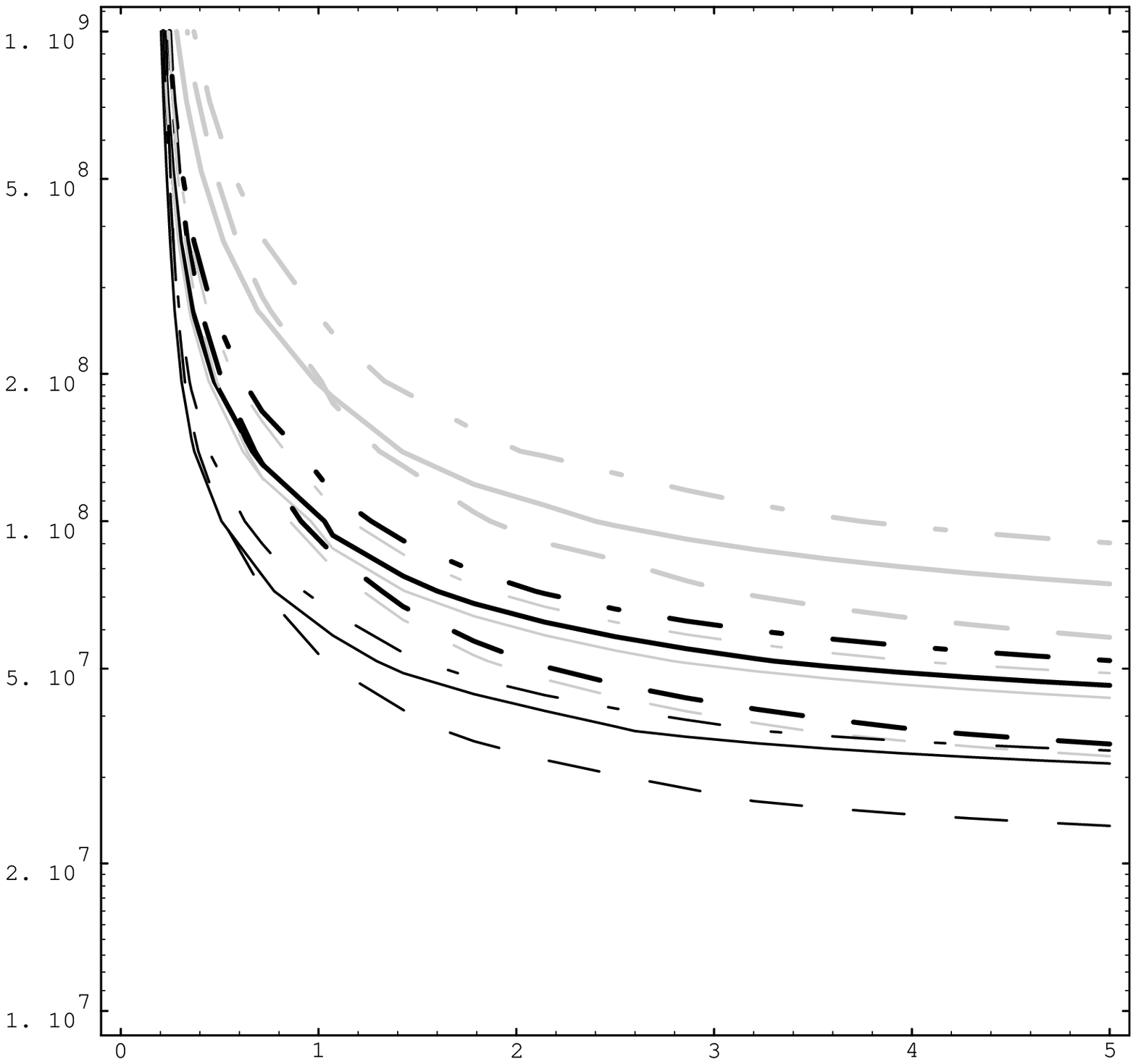}
\ec
\vskip -5.0in
{\hskip 3.2in \epsfxsize=2.8in \epsffile{fig07.ps}}
\vskip -0.3in
{\hskip -0.1in \Large \rotate[l]{$\sigrc$}}
\vskip 1.8in
{\hskip 3.1in \Large $z_{s}$}
\vskip 0.2in

\figcaption{Contour plots of lensing probability as for Fig. 1, but integrating
over a Schechter distribution of lenses}

\newpage
\bc\
\epsfxsize=5in
\epsffile{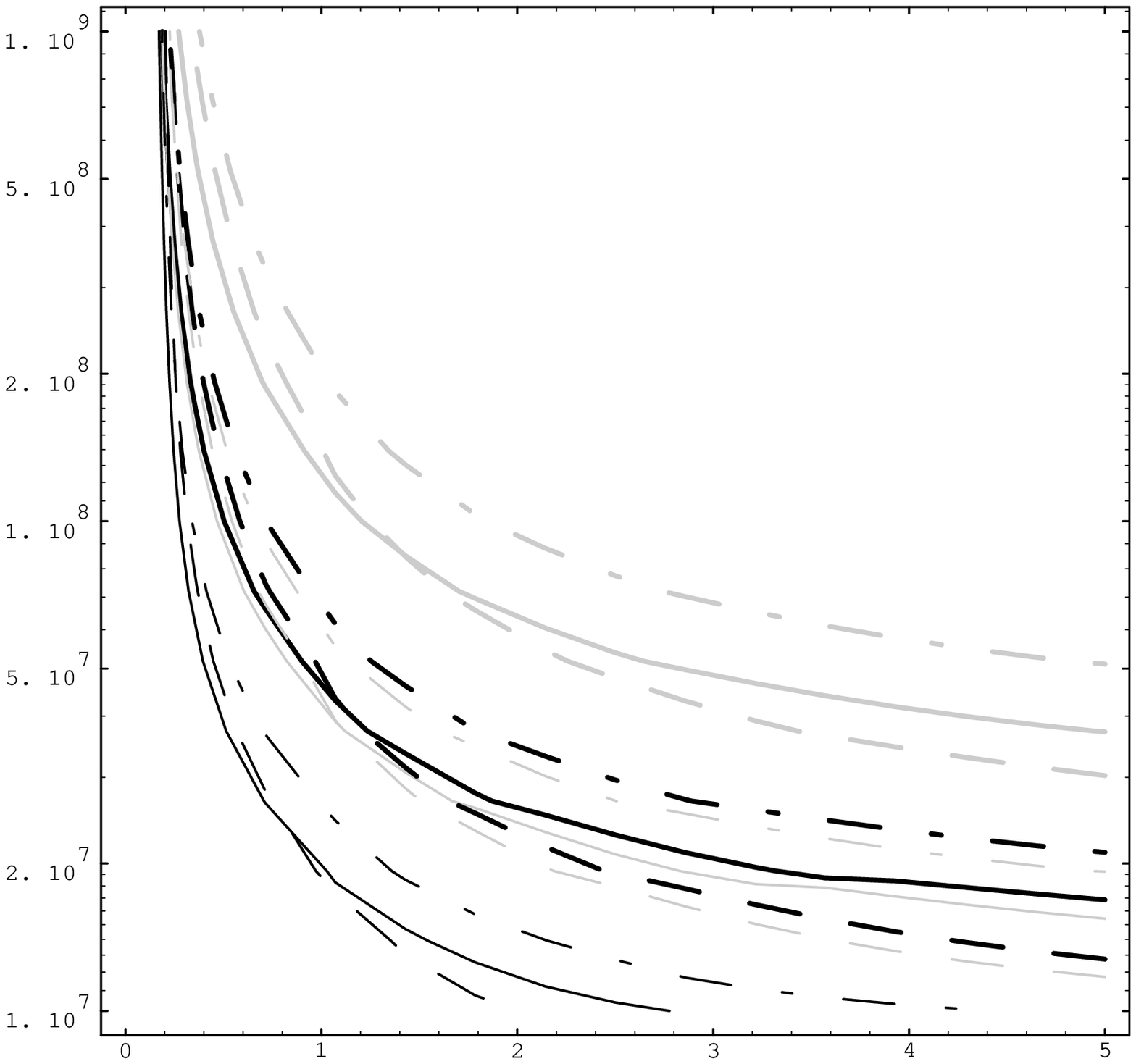}
\ec
\vskip -5.0in
{\hskip 2.8in \epsfxsize=3.2in \epsffile{fig07.ps}}
\vskip -0.3in
{\hskip -0.1in \Large \rotate[l]{$\sigrc$}}
\vskip 1.8in
{\hskip 3.1in \Large $z_{s}$}
\vskip 0.2in

\figcaption{As for Fig. 2, but with $\alpha_{2} = -1.5$. Note that all 
curves involve smaller value of $\sigrc$ compared to those in
Fig. 2.}

\newpage
\bc\
\epsfxsize=5in
\epsffile{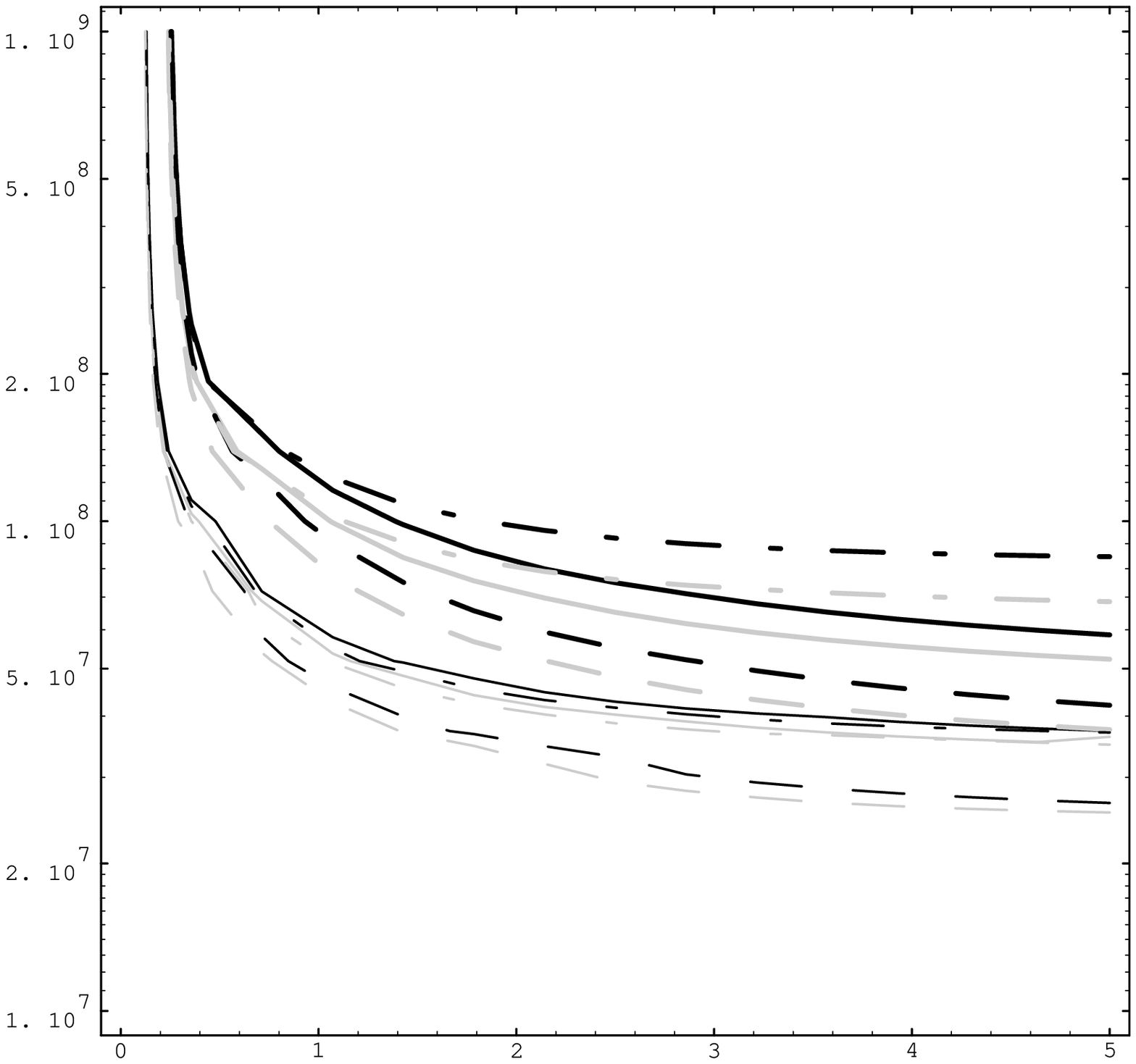}
\ec
\vskip -5.0in
{\hskip 2.6in \epsfxsize=3.0in \epsffile{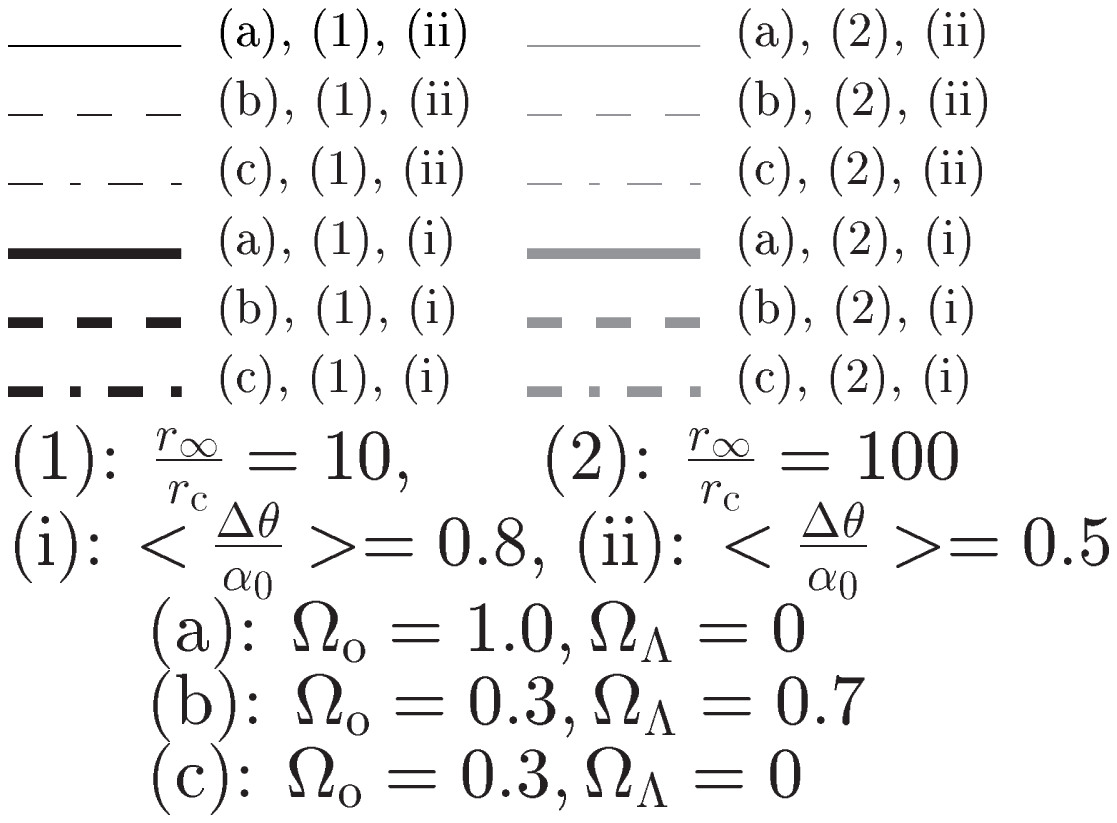}}
\vskip -0.3in
{\hskip -0.1in \Large \rotate[l]{$\frac{(\sigma/km/sec)^{2}}{\rc/(h^{-1}Mpc)}$}}
\vskip 1.8in
{\hskip 3.1in \Large $z_{s}$}
\vskip 0.2in

\figcaption{Contour plots of the expectation values of 
angular splitting normalized to that for an isothermal
sphere, for fixed lens mass and showing contours 
$< \dthefa > = 0.5$ and 0.8. The thicker width lines represent
$<\dthefa>=0.8$, and the thinner width lines represent $<\dthefa>=0.5$.
All other features are as for Fig.~1.}

\newpage
\bc\
\epsfxsize=5in
\epsffile{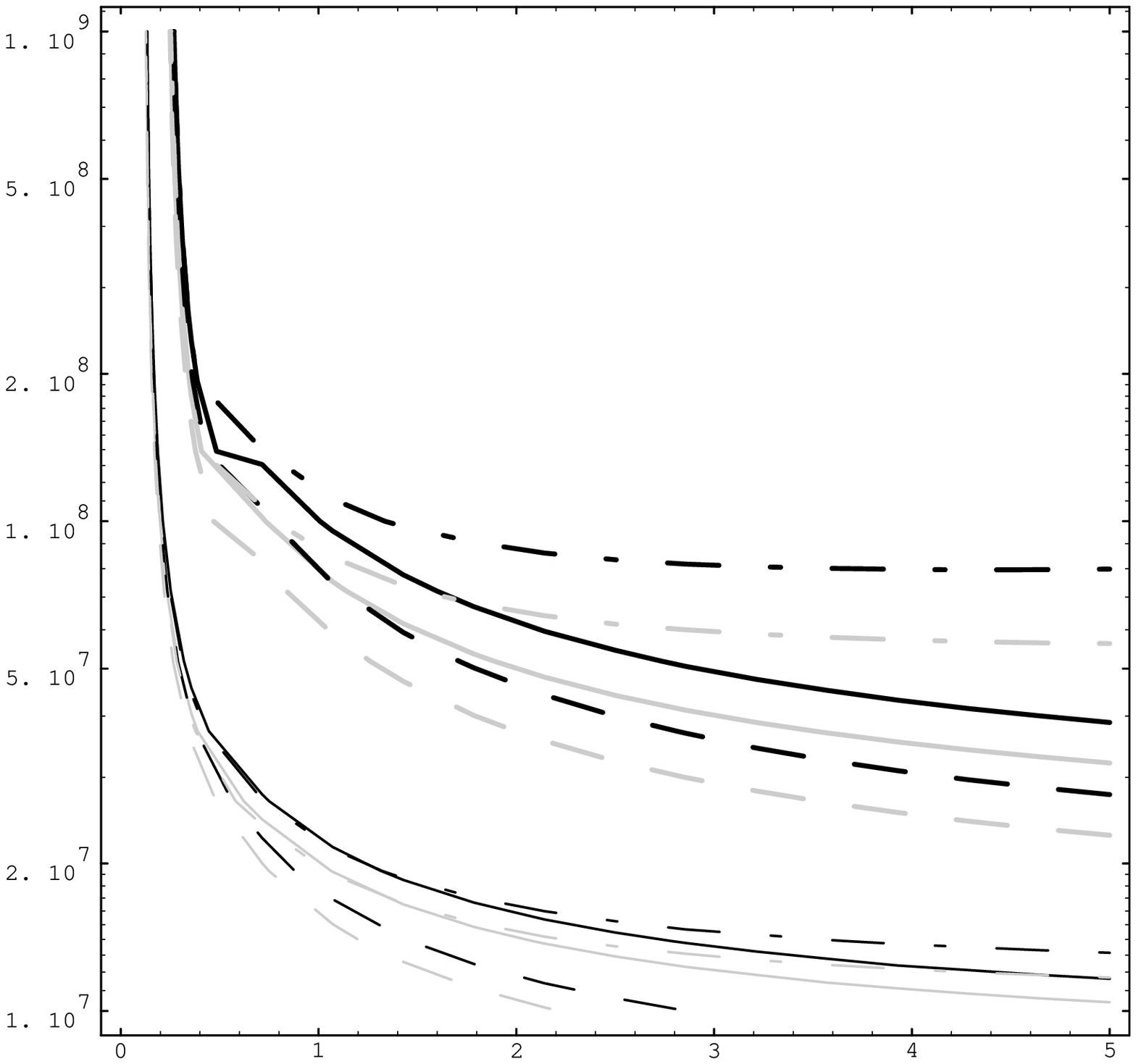}
\ec
\vskip -5.0in
{\hskip 2.6in \epsfxsize=3.0in \epsffile{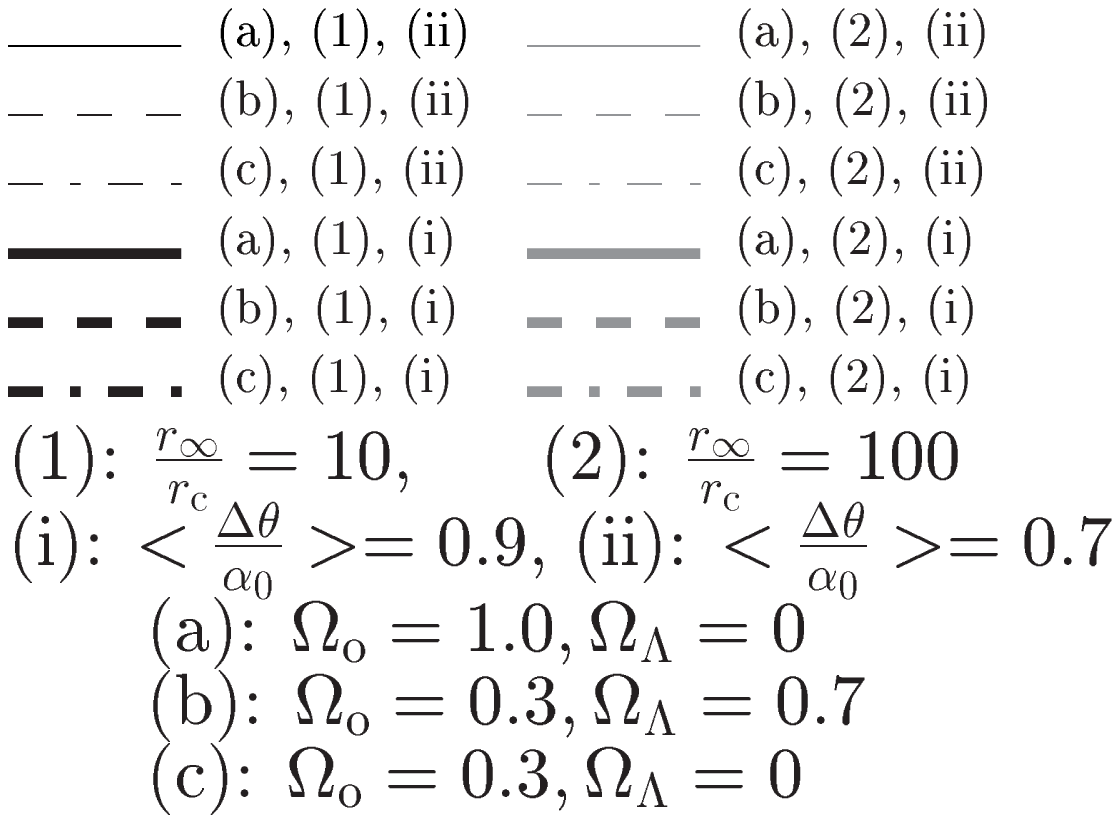}}
\vskip -0.3in
{\hskip -0.1in \Large \rotate[l]{$\sigrc$}}
\vskip 1.8in
{\hskip 3.1in \Large $z_{s}$}
\vskip 0.2in

\figcaption{As for Fig.~4, but integrating over a Schechter distribution,
with
$\alpha_{2} = -0.5$
and showing, $< \dthefa > =$ 0.7 and 0.9.}

\newpage
\bc\
\epsfxsize=5in
\epsffile{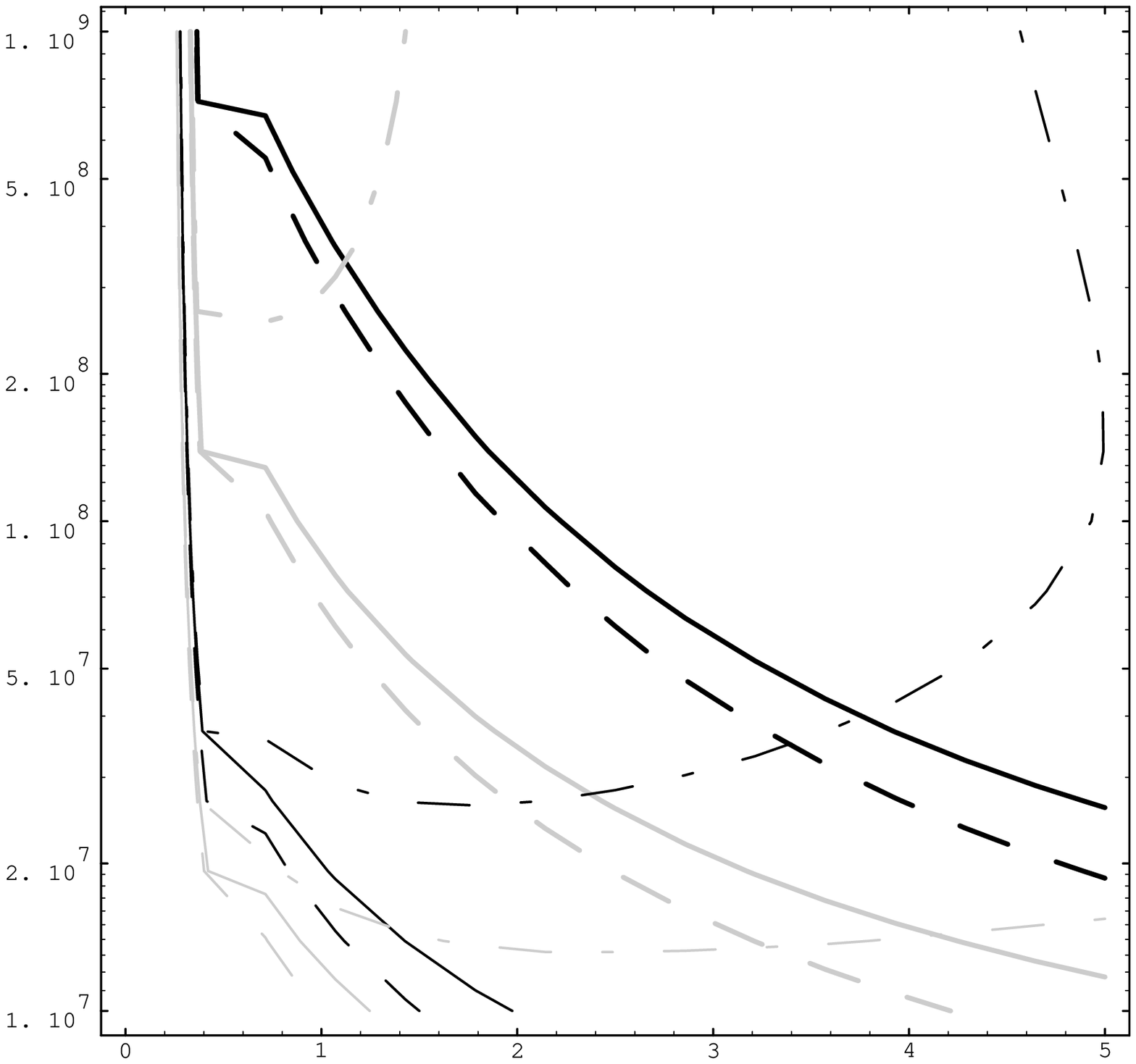}
\ec
\vskip -5.0in
{\hskip 2.65in \epsfxsize=2.7in \epsffile{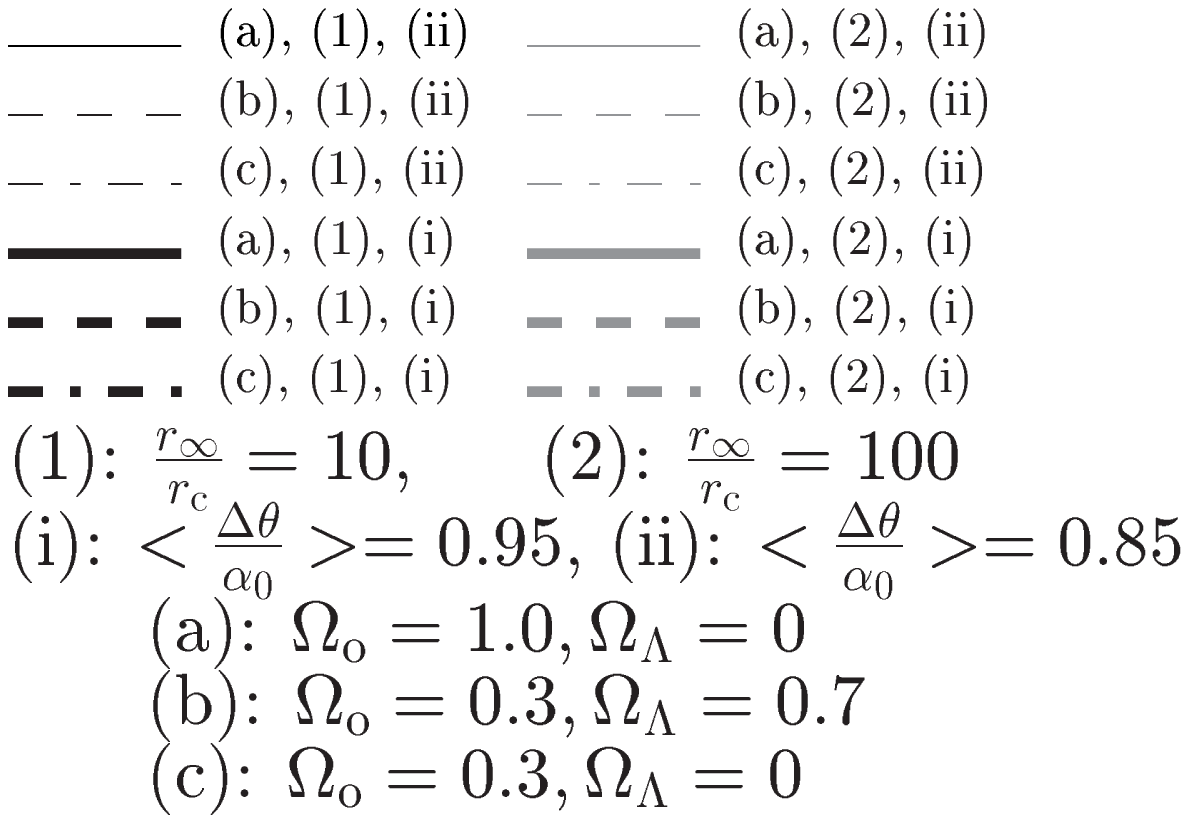}}
\vskip -0.3in
{\hskip -0.1in \Large \rotate[l]{$\sigrc$}}
\vskip 1.8in
{\hskip 3.1in \Large $z_{s}$}
\vskip 0.2in

\figcaption{As for Fig.~5, but for $\alpha_{2} = -1.5$
and showing, $< \dthefa > =$ 0.85 and 0.95. Note that all 
curves involve smaller values of $\sigrc$ compared to those in Fig. 5.}

\end{document}